\documentclass[fleqn,usenatbib]{mnras}

\usepackage{newtxtext,newtxmath}
\usepackage{tabularx}
\usepackage{comment}
\usepackage{physics}

\usepackage[T1]{fontenc}

\DeclareRobustCommand{\VAN}[3]{#2}
\let\VANthebibliography\thebibliography
\def\thebibliography{\DeclareRobustCommand{\VAN}[3]{##3}\VANthebibliography}

\usepackage{graphicx}	
\usepackage{amsmath}	
\usepackage{xcolor}
\usepackage{soul}
\usepackage{mathtools}

\newcommand{\qint}{\dot{Q}_{\text{dep}}^{\text{int}}}
\newcommand{\qdep}{\dot{Q}_{\text{dep}}}
\newcommand{\qinst}{\dot{Q}_{\text{inst}}}
\newcommand{\qdot}{\dot{Q}_{\beta}}
\title[Analytic Electron Deposition Formulae]{An Analytic Description of Electron Thermalization in Kilonovae Ejecta}

\author[B. Shenhar et al.]{
Ben Shenhar,\thanks{E-mail: ben.shenhar@weizmann.ac.il}
Or Guttman
and Eli Waxman
\\
Department of Particle Physics \& Astrophysics, Weizmann Institute of Science, Rehovot 76100, Israel
}

\date{Accepted XXX. Received YYY; in original form ZZZ}

\pubyear{2024}

\begin{document}
\label{firstpage}
\pagerange{\pageref{firstpage}--\pageref{lastpage}}
\maketitle

\begin{abstract}

A simple analytic description is provided of the rate of energy deposition by $\beta$-decay electrons in the homologously expanding radioactive plasma ejected in neutron star mergers, valid for a wide range of ejecta parameters- initial entropy, electron fraction \{$s_0,Y_e$\} and density $\rho t^3$. The formulae are derived using detailed numerical calculations following the time-dependent composition and $\beta$-decay emission spectra (including the effect of delayed deposition). The deposition efficiency depends mainly on $\rho t^3$ and only weakly on $\{s_0,Y_e\}$. The time $t_e$ at which the ratio between the rates of electron energy deposition and energy production drops to $1-e^{-1}$, is given by $t_e=t_{0e}\Big(\frac{\rho t^3}{0.5(\rho t^3)_0}\Big)^a$, where $(\rho t^3)_0=\frac{0.05M_{\odot}}{4\pi(0.2c)^3}$, $t_{0e}(s_0,Y_e)\approx17$~days and $0.4\le a(s_0,Y_e)\le0.5$. The fractional uncertainty in $t_e$ due to nuclear physics uncertainties is $\approx10\%$. The result $a\le0.5$ reflects the fact that the characteristic $\beta$-decay electron energies do not decrease with time (largely due to "inverted decay chains" in which a slowly-decaying isotope decays to a rapidly-decaying isotope with higher end-point energy). We provide an analytic approximation for the time-dependent electron energy deposition rate, reproducing the numerical results to better than 50\% (typically $<30\%$, well within the energy production rate uncertainty due to nuclear physics uncertainties) over a 3-4 orders-of-magnitude deposition rate decrease with time. Our results may be easily incorporated in calculations of kilonovae light curves (with general density and composition structures), eliminating the need to numerically follow the time-dependent electron spectra. Identifying $t_e$, e.g. in the bolometric light curve, will constrain the (properly averaged) ejecta $\rho t^3$.

\end{abstract}




\section{Introduction} \label{introduction}
The violent merger of two neutron stars (or a neutron star and black hole) has been predicted to produce high-density neutron-rich ejecta, in which heavy elements beyond Iron are formed via the $r$-process (\citet{lattimer_black-hole-neutron-star_1974}, see \citet{shibata_merger_2019,radice_dynamics_2020,rosswog_heavy_2024} for recent reviews). The radioactive decay of these freshly synthesized, unstable $r$-process nuclei should power an optical transient \citep{li_transient_1998} commonly referred to as a "kilonova" (see \citet{fernandez_electromagnetic_2016,nakar_electromagnetic_2020,margutti_first_2021} for recent reviews). The observed UV-IR emission following the neutron star merger event GW170817 \citep{arcavi_optical_2017, coulter_swope_2017} is broadly consistent with these predictions (\cite{drout_light_2017, hotokezaka_neutron_2018, kasen_origin_2017, kasliwal_illuminating_2017, waxman_constraints_2018}), confirming the existence of an $r$-process powered transient with $M_{ej} \approx 0.05 M_{\odot}$ and characteristic velocity $v_{ej} \approx 0.2 c$.  \citep{cowperthwaite_electromagnetic_2017,kasen_origin_2017, tanaka_kilonova_2017,tanvir_emergence_2017,rosswog_first_2018}. Nonetheless, an outstanding problem in kilonovae analysis is the translation of kilonovae observations to robust constraints on the ejecta properties, in particular on the $r$-process nucleosynthesis and on the final ejecta composition.

Interpreting kilonovae observations requires understanding the processes by which secondary particles produced in radioactive decay deposit energy, i.e., “thermalize,” in the ejected plasma. Radioactivity produces energetic particles (photons, electrons, alphas, and fission fragments). Initially, the particles' energy is wholly absorbed in the plasma. However, as the ejecta expand and dilute, the energy is only partially absorbed, leading to an energy deposition rate that deviates from the total energy release rate. $\gamma$-rays initially dominate the radioactive heating, up to $t \sim 1$ day at which time they escape the plasma \citep{barnes_radioactivity_2016, guttman_thermalization_2024}. On a timescale of $t \sim 1-10$ days, the main heating source is expected to be $\beta$-decay electrons, with possible $\alpha$ and fission heating contributions depending on initial $Y_e$ and nuclear mass model: For initial $Y_e \gtrsim 0.2$, $\beta$-decay electron heating is expected to dominate the heating for all $t \gtrsim 1$ day, with a possible comparable contribution from fission heating at $t \sim 100$ days for some nuclear mass models \citep[e.g][]{zhu_modeling_2021}; For initial $Y_e \lesssim 0.2$, mass model uncertainties yield larger variance in the heating contributions- $\alpha$-particle heating was found to be of the same order as $\beta$-decay electron heating for $t \lesssim 10$ days for some mass models \citep[e.g.][]{zhu_modeling_2021, barnes_kilonovae_2021} and fission fragments may be equally important on timescales of $10 \lesssim t \lesssim 100$ for most mass models investigated \citep[However, the fission contribution is highly uncertain due to uncertainties in theoretical mass models, fission barriers, and fission rates, e.g.][]{zhu_modeling_2021, barnes_kilonovae_2021}. Therefore, notwithstanding the effects of fission, late-time kilonovae light-curve interpretation requires an understanding of the inefficient thermalization timescale on which electrons efficiently deposit their energy, particularly for compositions with initial $Y_e \gtrsim 0.2$.  

In this work, we calculate the time-dependent energy deposition rate per unit mass, $\qdep$, by electrons and positrons (in what follows, we refer to both as "electrons") emitted by $\beta$-decays in homologously expanding plasma ejected in binary neutron star (BNS) mergers for a wide range of ejecta parameters, $\{\rho t^3,s_0, Y_e\}$. Using detailed numeric nucleosynthesis calculations, we obtain the time-dependent composition, stopping power, and $\beta$-decay energy spectra, and follow the evolution of the electron energy distribution, assuming that the electrons are confined by magnetic fields to the fluid element within which they were produced\footnote{ 
Magnetic confinement appears likely since it would be facilitated by a relatively weak magnetic field. Assuming $B\propto r^{-2}$ in the expanding ejecta, as may be appropriate for a tangled field, $B \sim 0.1 \mu$G is expected at $t \sim 10$ days, $r \sim 10^{16}$ cm (the equipartition field at 10 days is $\sim1$ mG), implying an electron Larmor radius of $\sim 10^{10}$ cm.}
(i.e., including "delayed energy deposition", the deposition at time $t$  by electrons that were emitted at $t'<t$ and did not lose all their energy by the time $t$). The secondary electrons that are ionized by $\beta$-decay electrons have very low kinetic energies and thermalize rapidly, we thus assume that they fully thermalize upon emission. 
Finally, we test the sensitivity of our results to uncertainties in nuclear physics.

We carry out calculations over a broad range of values of ejecta parameters $ \{ s_0, Y_e, \rho t^3\} $, that generously covers the expected ejecta parameter range from binary neutron star merger simulations \citep[e.g.][]{radice_binary_2018, nedora_numerical_2021} : $ 10^{-3} \leq \rho t^3/( \rho t^3)_0 \leq 10^2$, where $( \rho t^3)_0 = \frac{0.05 M_\odot}{4 \pi (0.2c)^3}$ is our benchmark value motivated by the inferred mass and characteristic velocity of the ejecta of GW170817 \citep{drout_light_2017, hotokezaka_neutron_2018,kasen_origin_2017, kasliwal_illuminating_2017, waxman_constraints_2018} (for homologous expansion, $\rho t^3$ is a constant of the ejecta determined by its mass and velocity spread); $1k_b/\text{baryon} \leq s_0\leq 100k_b/\text{baryon}$; $0.05 \leq Y_e\leq 0.45$. 
Our results may be readily applied to ejecta with general density and composition structures by properly integrating over the ejecta components. Under the assumption that electrons are confined to the plasma by magnetic fields, such integration would be valid also at late times, $>t_e$, at which delayed deposition may become important. 


We provide a simple analytic description of the thermalization efficiency that is valid for a wide range of ejecta parameters, \{$\rho t^3$, $s_0, Y_e$\}, extending the results of earlier works, that compile tables of parameterized fits to the thermalization efficiency for specific combinations of ejecta mass and velocity \citep[e.g.][]{barnes_radioactivity_2016}. 
Although our calculated $t_e$ values are broadly consistent with those of earlier works \citep{metzger_electromagnetic_2010, barnes_radioactivity_2016,hotokezaka_radioactive_2020, kasen_radioactive_2019}, our analytic approximation differs from earlier analytic results. This is partly due to our calibration against a systematic numeric analysis, and partly due to assumptions adopted in earlier work, which we find invalid (see \S~\ref{results section}). In particular, it is commonly assumed that the characteristic energy of electrons released in $\beta$-decays drops with time \citep[e.g.][]{kasen_radioactive_2019, hotokezaka_radioactive_2020}, leading to a dependence of $t_e$ on $\rho t^3$ which is stronger than the $t_e\propto(\rho t^3)^{1/2}$ scaling expected for time-independent characteristic energy (see discussion in \S~\ref{char energy release electrons}). While the $Q$-value of $\beta$-decays decreases with decay half-time $\tau$, this does not imply \citep[as pointed out in][]{waxman_late-time_2019} a decrease of the characteristic $\beta$-decay electron energy with time $t$, since there is a very large spread in the relation between $Q$ and $\tau$ and since the decay of an isotope with a low $Q$-value and a long lifetime may produce an unstable isotope with a short lifetime and a high $Q$-value. We find here that such "inverted decay chains" lead to a nearly time-independent, or slightly rising with time, characteristic $\beta$-decay electron energy, and $t_e\propto(\rho t^3)^{a}$ with $a\lesssim0.5$.

The plasma stopping power depends on the ejecta composition and on the energy spectrum of the $\beta$-decay electrons, which in turn depend on $Y_e$ (and $s_0$). $t_e$ is therefore often expected to depend strongly on $Y_e$ and nuclear physics inputs \citep{barnes_radioactivity_2016, barnes_kilonovae_2021, zhu_modeling_2021}. We find that the deposition efficiency, particularly $t_e$, depends primarily on $\rho t^3$, with only a weak dependence on $\{Y_e,s_0\}$ and on nuclear physics uncertainties. This is due to the fact that the ionization stopping power. which dominates the energy loss, depends on the composition mainly through the ratio of atomic number and mass, $Z/A$, which is nearly composition independent, and since the characteristic $\beta$-decay electron energy depends weakly on composition.

This paper is organized as follows. In \S~\ref{methods}, we describe our nucleosynthesis and electron spectra calculations. In \S~\ref{thermalization section}, we give the equations we used for (numerically) calculating the electron energy loss and deposition, and the definition of $t_e$. \S~\ref{results section} presents our results. In \S~\ref{analytical formulae section} we present our analytical formulae for $t_e$. In \S~\ref{char energy release electrons} we discuss the characteristic energy of $\beta$-decay electrons. In \S~\ref{qdep analytic description} we present our analytical formulae for the energy deposition rate $\dot{Q}^{int}_{dep}(t)$. \S~\ref{robustness checks} presents an analysis of the sensitivity of our results to nuclear physics uncertainties. We summarize and discuss our conclusions in  \S~\ref{conclusions}.

\section{Nucleosynthesis and Electron Spectra Calculation} \label{methods}

We carry out nucleosynthesis calculations over a $10\cross$12$\cross$9 grid of \{$\rho t^3$,$s_0,Y_e$\} values, with logarithmic spacing in $ 10^{-3} \leq \rho t^3/( \rho t^3)_0 \leq 10^2$, semi-linear spacing in $1 \leq s_0/(k_b/\text{baryon})\leq 100$, and linear spacing in $0.05 \leq Y_e\leq 0.45$. BNS merger simulations show that most dynamical ejecta and secular wind ejecta have an entropy $s_0 \approx 20 k_b/\text{baryon}$, while a small mass fraction of the dynamical ejecta ($M<10^{-1} M_{ej}$) has $s_0 \geq 50 k_b/\text{baryon}$ \citep{radice_binary_2018,nedora_numerical_2021}.

We compute time-dependent yield abundances using the publicly available \textit{SkyNet} nuclear network, including 7843 isotopes up to $^{337}\textrm{Cn}$ \citep{lippuner_skynet_2017}. \textit{SkyNet} requires time-dependent trajectories of Lagrangian fluid elements to predict the temporal evolution of the abundances. Matter density evolves first through an exponential phase which is largely constant, before transitioning to a homologous expansion \citep{lippuner_r-process_2015}:

\begin{equation} \label{skynet density history}
    \rho(t) = 
    \begin{cases}
    \rho_0 e^{-t/\tau} & \text{for } t\leq 3 \tau \\
    \rho_0 \Big( \frac{3 \tau}{et} \Big)^3,              & \text{otherwise}
\end{cases}
\end{equation}
where $\tau$ is the expansion timescale of the ejecta. NSE is assumed to hold until $T=7 \text{GK}$, at which time \textit{SkyNet} switches to a full network evolution. Most works that examine \textit{SkyNet} nucleosynthesis yields initialize the calculation with given $\tau, T_0, Y_e, s_0$ \citep{perego_production_2022, lippuner_r-process_2015}, where $T_0$ is the initial temperature of the ejecta. For these given values, the EoS determines the initial density $\rho_0$ (see \citet{lippuner_skynet_2017} regarding EoS and NSE solver used by \textit{SkyNet}) and then uses the density history provided by Eq. (\ref{skynet density history}) to evolve the network. 

In contrast, we initialize our nucleosynthesis calculations as follows: All runs are initialized with $T_0=10$ GK, as NSE is expected for this temperature. We supply initial values of $s_0, Y_e$, and then find $\rho_0$ using \text{SkyNet}'s EoS solver. We determine $\tau$ from $\rho_0$ using Eq. (\ref{skynet density history}), such that the ejecta has the desired $\rho t^3$ value. We stress that in this approach $\tau$ is not a meaningful physical parameter; we assume that the ejecta has a particular $\rho t^3$, $s_0$ and was initially in NSE, in which case $\tau$ specifies the time before expansion. Initializing with different $T_0$ would result in a different $\tau$, however as NSE conditions hold for both, the same composition is achieved. This alternative approach allows us to directly probe the dependence on $\rho t^3$ for fixed $s_0$ within the range obtained in BNS simulations.

We employ the latest JINA REACLIB database \citep{cyburt_jina_2010}, with specific corrections as described in Appendix B of \citet{guttman_thermalization_2024}, using the same setup as in \citet{lippuner_r-process_2015} for the other input nuclear physics. More specifically, strong inverse rates are computed assuming detailed balance. Spontaneous and neutron-induced fission rates are taken from \citet{frankel_calculations_1947, panov_dynamics_2009}, adopting fission barriers and fission fragment distributions from \citet{mamdouh_fission_2001, wahl_systematics_2002}. Nuclear masses are taken from the REACLIB database, which includes experimental masses and theoretical masses calculated from the FRDM model \citep{moller_nuclear_2016}.

We use experimental data from the latest ENDF database \citep{brown_endfb-viii0_2018} to calculate the time-dependent energy released in electrons. We ensure that the total energy release we compute is equivalent to $\dot{M} c^2$ as calculated by \textit{SkyNet}. We use \textit{BetaShape} \citep{mougeot_betashape_2017} to calculate spectra of individual $\beta$-decays. We then compute the full time-dependent emission spectra using the calculated activities of $\beta$-decaying isotopes.

\section{Radioactive Heating} \label{thermalization section}

In this section, we first present the energy-loss mechanisms for electrons, and then describe the energy deposition calculation method and the definition of $t_e$.

\subsection{Electron energy loss \label{radiactive heating losses}}

$\beta$-decay electrons emitted in the $10^{-1}-1$ MeV range lose energy primarily through ionization losses. However, for initial entropies $s_0 \approx 10^2 k_b / \text{baryon}$, the ejecta composition is dominated by H and/or He \citep{perego_production_2022}. In this case, plasma losses dominate over other energy loss mechanisms. 

We calculate mass-weighted, time-dependent stopping powers based on the instantaneous composition of the ejecta,
\begin{equation}
    \Big( \frac{dE}{dX} \Big)_{\rm tot} = \sum_{\rm iso} A_{\rm iso}Y_{\rm iso} \Big( \frac{dE}{dX} \Big)_{\rm iso}.
\end{equation}
Where $\Big( \frac{dE}{dX} \Big)_{\rm iso}$ is the energy loss per unit column density for a particular isotope, $A_{\rm iso}$ is the isotope's mass number and $Y_{\rm iso}$ is its number fraction, defined such that $\sum_{\rm iso} Y_{\rm iso} = 1$.
For electron ionization losses, we use \citep{longair_high_2011}
\begin{equation} \label{electron ionization losses}
\begin{split}
    \Big( \frac{dE_e}{dX} \Big) _{\text{ion}} & = \frac{4 \pi e^4}{m_e m_p v_e^2}\frac{Z}{A} \Bigg[ \text{ln}\Big( \frac{\gamma_e^2 m_e v_e^2}{\bar{I}} \Big) - \frac{1}{2} \text{ln}(1+\gamma_e) \\
   & - \Big( \frac{2 \gamma_e + \gamma_e^2 -1}{2 \gamma_e^2} \Big) \text{ln}2 + \frac{1}{2 \gamma_e^2} + \frac{1}{16} \Big( 1- \frac{1}{\gamma_e} \Big)^2 \Bigg],
\end{split}
\end{equation}
where $dE_e/dX$ is the energy loss per unit column density traversed by the electron, $dX = \rho dx$, $v_e$ is the electron's velocity, $\gamma_e$ is the Lorentz factor of the electron, $Z$ and $A$ are the atomic and mass numbers of the plasma nuclei, and $\bar{I}$ is the effective average ionization potential which can be approximated for an element of atomic number Z as \citep{emilio_segre_nuclei_1977}: $\bar{I} = 9.1Z \Big( 1 + \frac{1.9}{Z^{2/3}} \Big) $ eV.

For plasma losses, we use \citep{solodov_stopping_2008}
\begin{equation} \label{electron plasma losses}
\begin{split}
    \Big( \frac{dE_e}{dX} \Big) _{\text{plasma}} & = \frac{2 \pi e^4}{m_e m_p v_e^2}\frac{\chi_e}{A} \Bigg\{ \text{ln}\Bigg[ \Big( \frac{E_e}{\hbar \omega_p} \Big)^2 \frac{\gamma_e+1}{2 \gamma_e^2} \Bigg] + 1 \\
    & + \frac{1}{8} \Big( \frac{\gamma_e-1}{\gamma_e} \Big)^2  - \Big( \frac{2 \gamma_e -1}{ \gamma_e^2} \Big) \text{ln}2 \Bigg\},
\end{split}
\end{equation}
where $\chi_e$ is the number of free electrons per atom, $\omega_p = (4\pi e^2 n_e/m_e)^{1/2}$ is the plasma frequency, and $n_e = \chi_e \rho / (A m_p)$ is the number density of free electrons. We use $\chi_e=1$ and $\hbar \omega_p = 10^{-7}$ eV (we keep $\omega_p$ fixed at this value since it affects the stopping power only logarithmically).

Finally, at highly relativistic energies, Bremsstrahlung losses significantly contribute to electron losses \citep{longair_high_2011}
\begin{equation} \label{electron brem losses}
\Big( \frac{dE_e}{dX} \Big) _{\text{brem}} = \frac{4 e^4}{m_e m_p v_e c} \frac{Z^2 e^2}{A \hbar c} \frac{E_e}{m_e c^2} \Bigg[ \text{ln} \Big( \frac{183}{Z^{1/2}} \Big) + \frac{1}{8} \Bigg].
\end{equation}

\begin{figure} 	\includegraphics[width=\columnwidth]{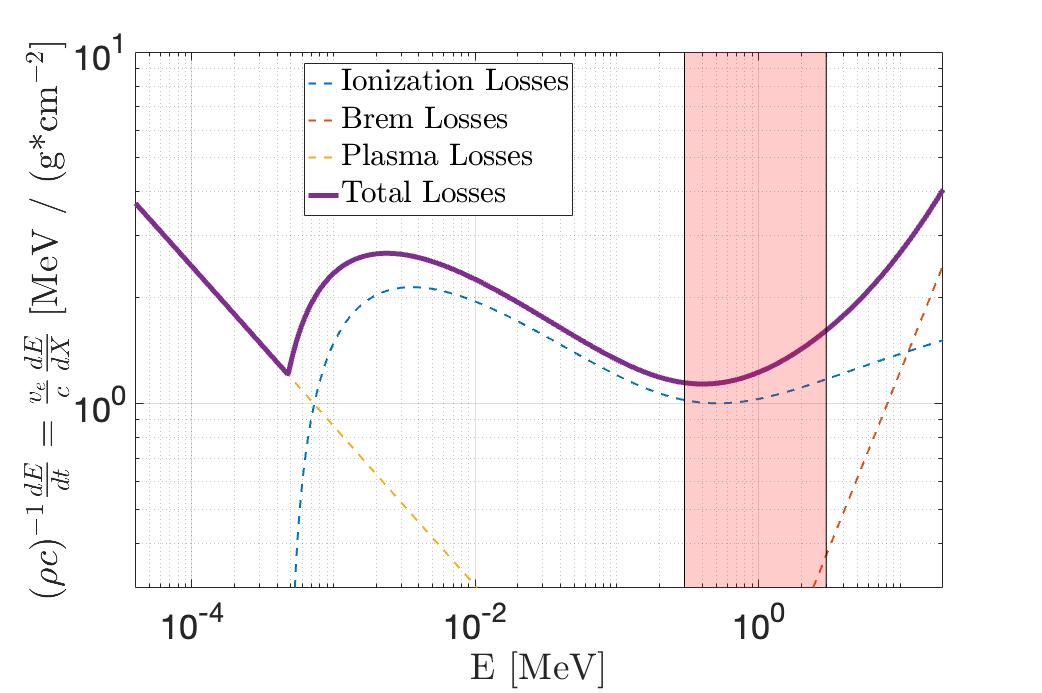}
    \caption{ Energy loss rate of electrons propagating in a singly ionized $\chi_e=1$ Xe plasma ($Z=54$, $A=131$) and $\hbar \omega_p = 10^{-7}$~eV. The shaded area shows typical average initial energies of $\beta$-decay electrons. For most relevant energies, ionization losses dominate.}
    \label{e_stopping_power_plot}
\end{figure}

Figure \ref{e_stopping_power_plot} shows the electron stopping power contributions for Xe composition. 

\begin{figure} 
	\includegraphics[width=\columnwidth]{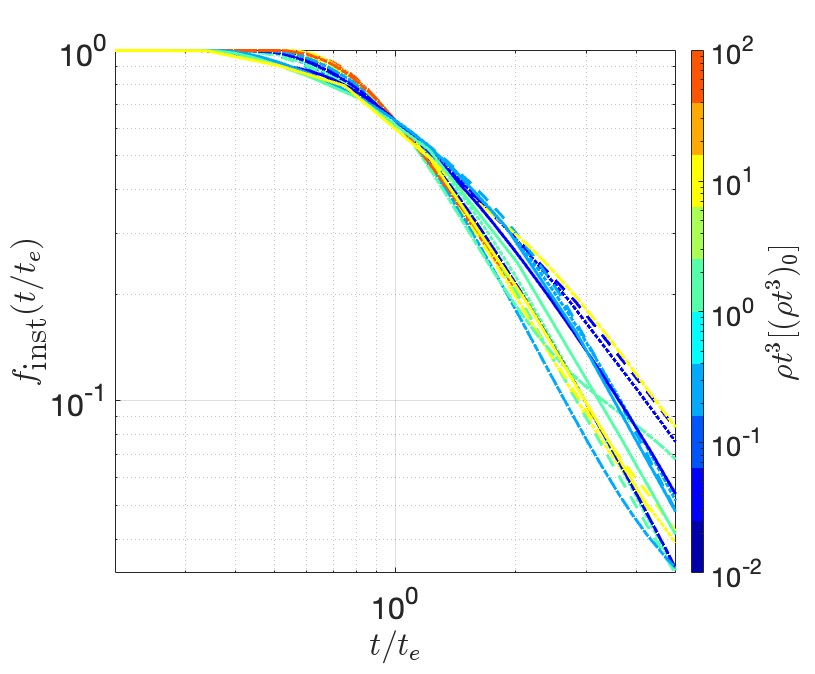}
    \caption{Instantaneous deposition fraction $f_{\text{inst}} (t/t_e)$ as defined in Eq. (\ref{f_dep tot}) for different ejecta parameters. Different colours correspond to different values of $\rho t^3$ and different line-styles correspond to different $Y_e$ values. Dotted, dashed, dashed-dotted, and solid lines correspond to $Y_e = 0.05, 0.15, 0.25, 0.4$, respectively. $s_0=20$ $k_b/\text{baryon}$ for all calculations shown. The instantaneous deposition fractions follow similar trajectories, especially before $t_e$.}
    \label{f_dep figure}
\end{figure}

\begin{figure*} 
	\includegraphics[width=\textwidth]{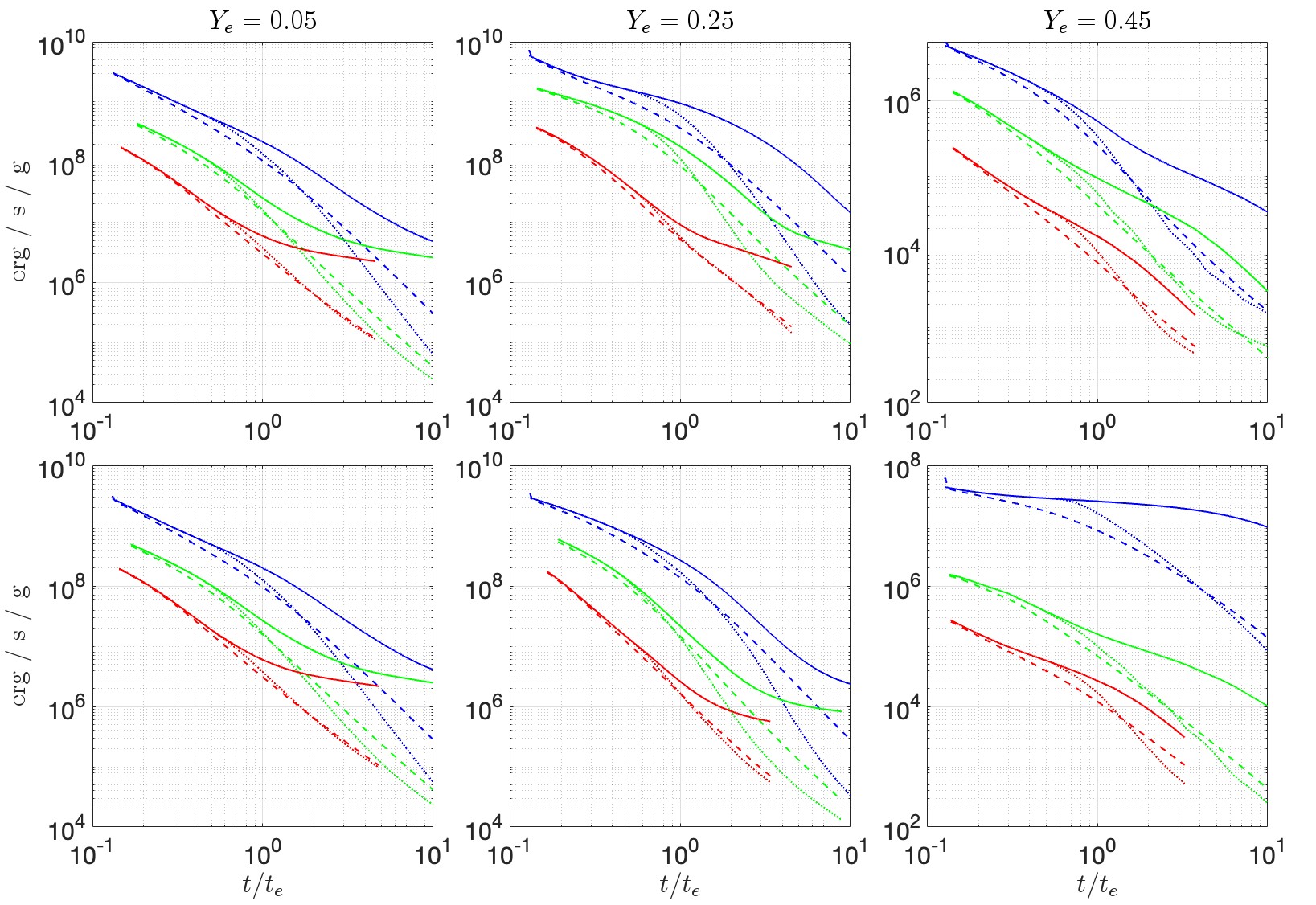}
    \caption{Electron energy release and energy deposition vs. $t/t_e$ for a selected sample of ejecta parameters. 
    Solid, dashed, and dotted lines show electron energy release $\dot{Q}_\beta$, full deposition calculation $\dot{Q}_{\text{dep}}$, and approximate instantaneous deposition $\dot{Q}_{\text{inst}} \equiv f_{\text{inst}} \dot{Q}$, respectively. $s_0=20, (60)$ $k_b/\text{baryon}$ for top (bottom) row. Left, middle, and right panels correspond to $Y_e = 0.05, 0.2, 0.45$, respectively. Blue, green, and red lines correspond to $\frac{\rho t^3}{(\rho t^3)_0} = 10^{-1}, 1, 10$ respectively. $\dot{Q}_{\text{dep}} < \dot{Q}_{\text{inst}}$ at $t<t_e$ due to adiabatic losses. At $t \sim 2 t_e$, $\dot{Q}_{\text{dep}} > \dot{Q}_{\text{inst}}$ as delayed deposition effects start to dominate. At late-times, $\dot{Q}_{\text{dep}} \sim t^{-2.85}$. }
    \label{comparing energy release}
\end{figure*}

\subsection{Electron Energy Deposition}

\subsubsection{Deposition Equations}

The energy deposited by electrons at time $t$ is given by
\newcommand{\dnde}{\frac{d N(E,t)}{dE}}
\begin{equation}
    \dot{Q}_{\text{dep}}(t) = \int dE\dnde \frac{dE}{dt} = \int dE \dnde \rho v_e \frac{dE}{dX},  
    \label{Q_dep def}
\end{equation}
where $\dnde$ is the number of electrons per unit energy, determined by the continuity equation in energy space
\begin{equation}
    \frac{\partial}{\partial t} \Big( \dnde \Big) = \frac{d \dot{N}}{dE} - \frac{\partial}{\partial E} \Big(\dot{E} \dnde \Big).  
    \label{dnde cont equation}
\end{equation}
Here, $\frac{d \dot{N}}{dE}$ is the number of electrons released per time per unit energy (as determined from the time-dependent abundances), and $\dot{E}$ is the full energy-loss rate for electrons, given by
\begin{equation}
    \dot{E} = -x \frac{E}{t} - \rho v \frac{dE}{dX}. \label{electron energy evolution}
\end{equation}
The first term on the right-hand side accounts for adiabatic energy losses and the second term accounts for energy transferred to the plasma as described in  \S~\ref{radiactive heating losses}. The adiabatic losses are obtained assuming that the high-energy electrons behave as an ideal gas, in which case we have $x \approx 1$ for highly relativistic electrons and $x \approx 2$ in the highly non-relativistic limit. $\beta$-decay electrons are mildly relativistic, with $\gamma_e v_e / c \sim 1$, and as they lose energy, the value of $x$ that best describes the evolution is changing. We take a fixed $x = 1$ throughout all our calculations. We numerically solve  Eq. (\ref{dnde cont equation}) for $\dnde$ and use the result in order to integrate Eq. (\ref{Q_dep def}). 

\subsubsection{Instantaneous Deposition and $t_e$}

We define the fraction of energy instantaneously deposited at time $t$ as
\begin{equation} \label{f_dep tot}
f_{\text{inst}}(t) = \frac{\dot{Q}_\text{inst}}{\dot{Q}_{\beta }} =
\frac{1}{\dot{Q}_{\beta}} \int dE f_{e}(E,t) E \frac{d \dot{N}(E,t)}{dE}, 
\end{equation}
where $\dot{Q}_{\beta}$ is the energy production rate in $\beta$-decay electrons, and the fraction of energy instantaneously deposited by an electron produced at time $t$ with energy $E_i$ is approximated as
\begin{equation} \label{f_dep}
    f_{e}(E_i, t) = 
    \begin{cases}
    1 & \text{for } t_l \leq t \\
    \frac{t}{t_l}   & \text{for } t_l \geq t .
\end{cases}
\end{equation}
Here $t_l(E_i,t) = E_i \Big( \frac{dE}{dt} \Big)^{-1} $ is the deposition energy loss timescale. Under this approximation, deposition deviates from full efficiency when the deposition energy loss timescale becomes larger than the dynamical time $t$. Note that at late times $f_e(E_i, t) \propto t^{-2}$ because $t_l \propto \rho^{-1} \propto t^{3}$. $t_e$ is defined as
\begin{equation} \label{t_e def}
f_{\text{inst}}(t_{e }) = 1 - e^{-1}.
\end{equation}

In Figure \ref{f_dep figure} we plot $f_{\text{inst}}(t/t_{e })$ for different ejecta parameters. 
At late times $f_{\text{inst}}(t)$ approaches a $1/t^{2}$ behavior, as expected from Eq. (\ref{f_dep}), although deviations from this behavior may occur due to variations in the emitted electron spectra. 
The rather common behavior of instantaneous deposition for different ejecta parameters suggests that an approximate analytic description may be obtained. 


In Figure \ref{comparing energy release}, we compare the rate of electron energy release and deposition. At early times, when $t<t_e$, $\dot{Q}_{\text{dep}} < \dot{Q}_{\text{inst}}$. This is due to energy that is lost to adiabatic expansion (first term on right-hand side of Eq. (\ref{electron energy evolution})). At $t = t_e$, $\qinst \simeq 1.5 \qdep$, with the exact factor depending on ejecta parameters. We are therefore confident in using our definition of $t_e$ as the inefficient deposition timescale, even though it is defined with respect to instantaneous deposition. At $t \sim 2 t_e$, $\dot{Q}_{\text{dep}} > \dot{Q}_{\text{inst}}$ as delayed deposition starts to dominate. At $t \gg t_e$, $\dot{Q}_{\text{dep}} \sim t^{-2.85}$ regardless of $\qdot (t)$, as was obtained analytically in \citet{waxman_late-time_2019}.

\section{Analytic approximations and nuclear physics uncertainties} \label{results section}

\subsection{Analytical approximation for $t_e$} \label{analytical formulae section}

\begin{figure} 
	\includegraphics[width=\columnwidth]{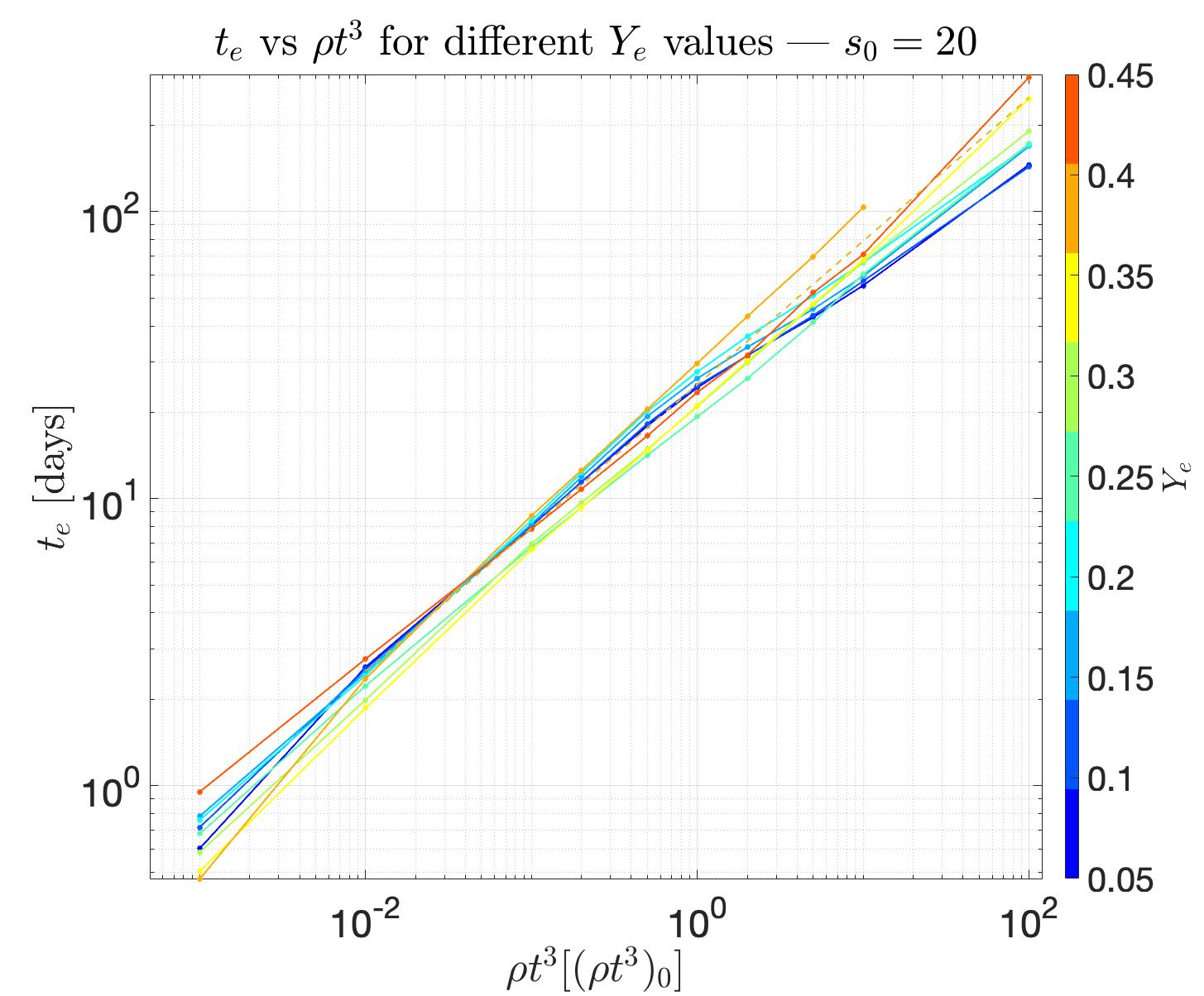}
    \caption{ $t_e$ vs. $\rho t^3$ for different $Y_e$ values and $s_0=20$ $k_b/\text{baryon}$. Grey dotted line shows a $t_e\propto(\rho t^3)^{1/2}$ fit. $t_e$ depends primarily on $\rho t^3$, with minor dependence on $Y_e$.}
    \label{t_e first plot}
\end{figure}

In Figure \ref{t_e first plot}, we plot $t_e$ vs. $\rho t^3$ for a range of $Y_e$ values (and $s_0=20$ $k_b/\text{baryon}$), together with a $t_e\propto (\rho t^3)^{1/2}$ fit. It is clear that the $t_e$ depends primarily on $\rho t^3$ and that a power-law fit is appropriate. However, there is a secondary $Y_e$ dependence. Particularly, for $Y_e<0.25$, $t_e$ drops below the $\sim(\rho t^3)^{1/2}$ curve for $\rho t^3 > 0.5(\rho t^3)_0$. We find that $\rho t^3 = 0.5(\rho t^3)_0$ serves as the breakpoint for a broken power-law description of $t_e$ for different combinations of $s_0$ and $Y_e$. Therefore, we suggest the following broken-power law description
\begin{equation} \label{t_e power law}
    t_{e} = 
    t_{0e} \begin{cases}
     \Big( \frac{\rho t^3}{0.5(\rho t^3)_0} \Big)^{a_1} \text{days} & \text{for } \rho t^3 < 0.5(\rho t^3)_0, \\
     \Big( \frac{\rho t^3}{0.5(\rho t^3)_0} \Big)^{a_2} \text{days} & \text{for } \rho t^3 > 0.5(\rho t^3)_0.
		 \end{cases}
\end{equation}

\begin{figure} 
	\includegraphics[width=\columnwidth]{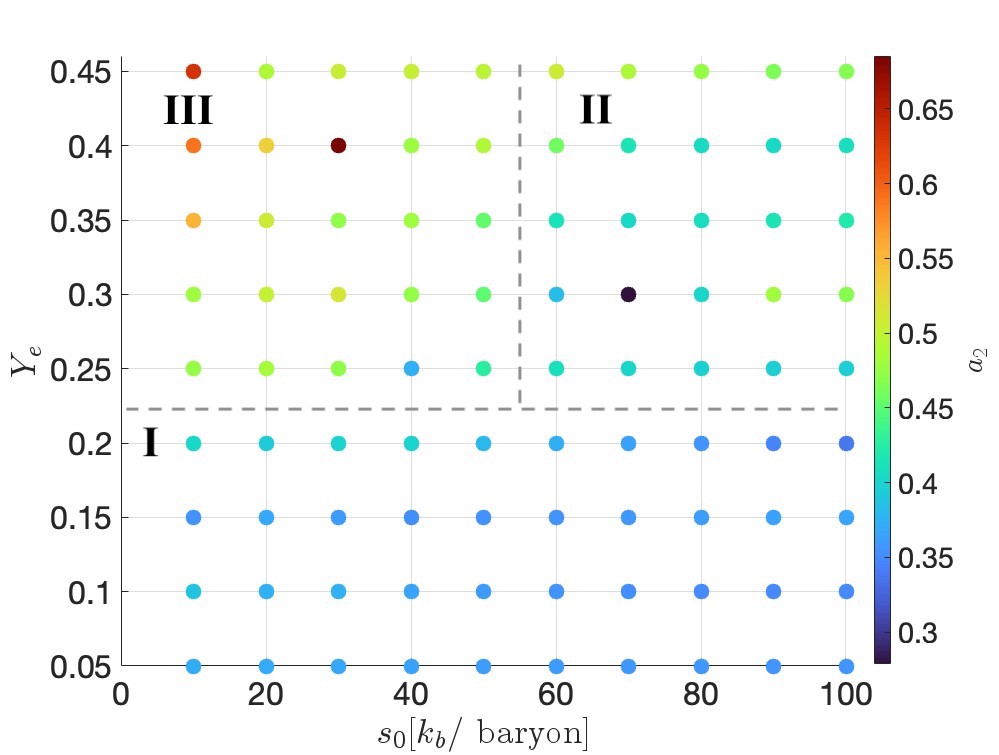}
    \caption{The exponent $a_2$ as defined in Eq. (\ref{t_e power law}) for different regions in the parameter space of \{$Y_e, s_0$\}. We identify 3 regions within which similar values of $a_2$ are obtained: Region I (bottom) - $Y_e<0.22$'; Region II (top-right) - $Y_e>0.22$ \& $s_0>55$ $k_b/\text{baryon}$; Region III (top-left) - $Y_e>0.22$  \& $s_0<55$ $k_b/\text{baryon}$.}
    \label{s2 plot}
\end{figure}

In Figure \ref{s2 plot}, we plot $a_2$ (as obtained from a fit of the numeric result to Eq. (\ref{t_e power law}) as a function of $s_0$ and $Y_e$. We identify 3 regions within which similar values of $a_2$ are obtained: Region I - $Y_e<0.22$; Region II - $Y_e>0.22$ \& $s_0>55$ $k_b/\text{baryon}$; Region III - $Y_e>0.22$ \& $s_0<55$ $k_b/\text{baryon}$. 
Nucleosynthesis calculations predict the production of 3rd-peak r-process elements in Region I \citep[e.g][]{perego_production_2022}, post 1st-peak elements in Region II (with 2nd-peak elements created for $Y_e = 0.25, 0.3$), and elements up to $A<125$ in Region III.

\newcommand{\rhoO}{ \frac{\rho t^3}{(\rho t^3)_0}}

\begin{figure} 
	\includegraphics[width=\columnwidth]{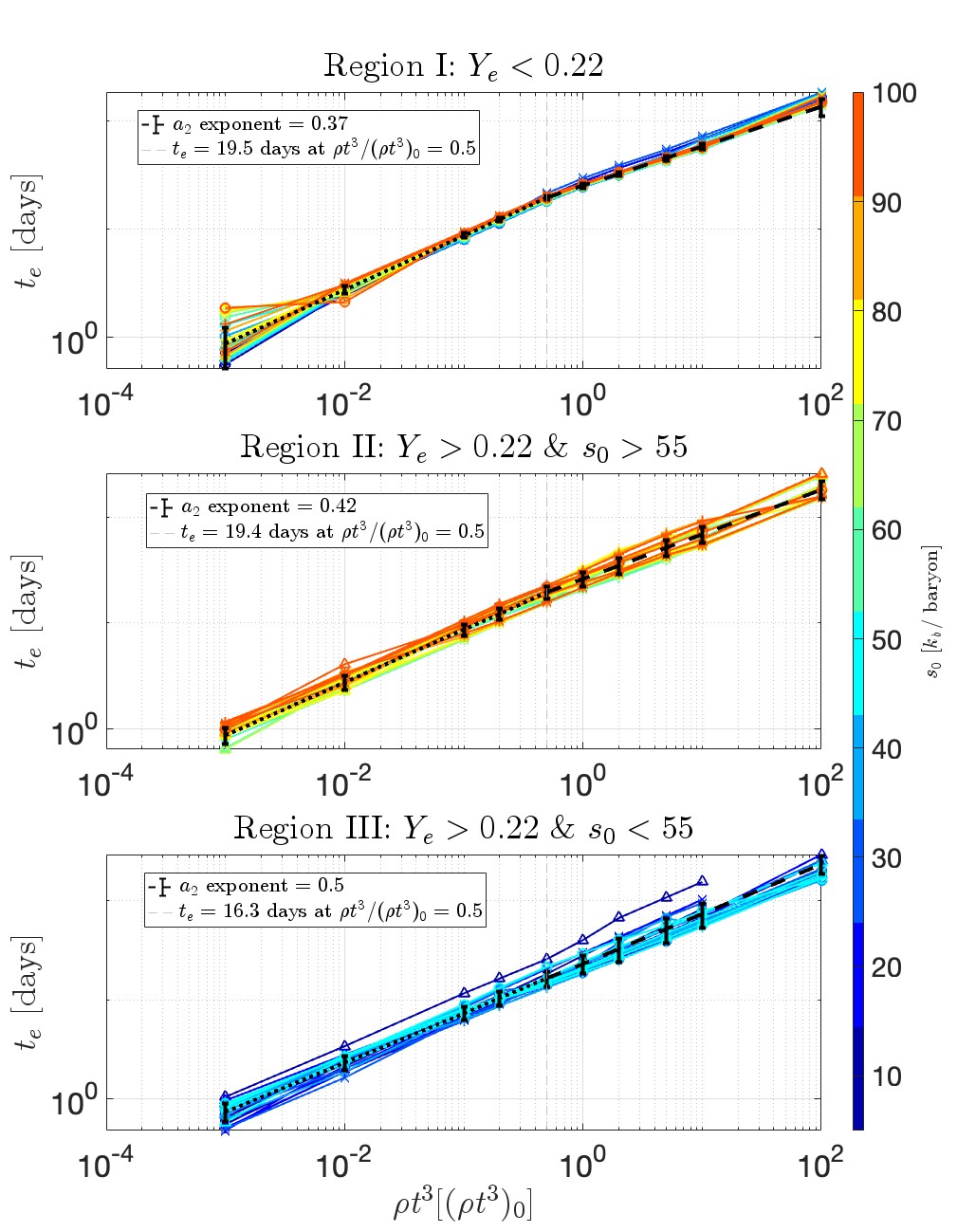}
    \caption{$t_e(\rho t^3)$ for the 3 regions in parameter space as defined in Figure \ref{s2 plot}, together with broken power-law fits as defined in Eq. (\ref{t_e power law}) and Table \ref{t_e table}. Different colors correspond to different $s_0$ values and different markers correspond to different $Y_e$ values. Error-bars show the standard-error of the estimate.}
    \label{t_e 3 regions plot}
\end{figure}

Table \ref{t_e table} presents the best-fit parameters for Eq. (\ref{t_e power law}) for the 3 regions considered. In Figure \ref{t_e 3 regions plot} we plot $t_e(\rho t^3)$ for all three regions, together with their respective broken power-law fits from Table \ref{t_e table}. Error bars show the standard error of the estimate, defined as $\sigma_e =  \sqrt{\frac{\sum_i (t_e-t_{e,i})^2}{N}}$ where $t_e$ is the value of the fit, calculated according to Eq. (\ref{t_e power law}), and $t_{e,i}$ is calculated according to Eq. (\ref{t_e def}) for the different $Y_e$ and $s_0$ values considered. The fit for Region I exhibits the best uniformity, with a fractional standard error (defined as $s_e \equiv \frac{\sigma_e}{t_e}$) of $s_e < 0.1$ for all $\rho t^3$ values excluding $\rhoO = 10^{-3}$, for which $s_e \approx 0.2$. For Regions II and III, $s_e \approx 0.1-0.2$, with the exact value depending on the value of $\rho t^3$.

We find that $a_1=1/2$ for our entire parameter space, while $a_2= 0.37, 0.42, 0.5$ for Regions I, II and III, respectively. As noted in the introduction, the fact that the power-law indices $a_{1,2}$ are close to $1/2$ implies that the characteristic energy of the $\beta$-decay electrons does not vary significantly with time, and the lower than $1/2$ values suggest a minor increase in this energy with time. We discuss this further in \S~\ref{char energy release electrons}.

\subsection{The characteristic energy of $\beta$-decay electrons}
\label{char energy release electrons}

The relation between the dependence of $t_e$ on $\rho t^3$ and the dependence of the characteristic $\beta$-decay electron energy on time may be qualitatively understood as follows. $t_e$ is approximately the time at which the expansion time, $t$, equals the deposition (mainly by ionization) energy loss time, $t_l=E_i(dE/dt)^{-1}$ where $E_i$ is the initial electron energy. Noting that $v_e dE/dX$ is only weakly dependent on energy at the relevant energy range (see Figure \ref{e_stopping_power_plot}), 
\begin{equation}
    \label{eq:tLoss}
    t_l=\frac{E_i}{dE/dt}=\frac{E_i}{\rho(v_e/c)dE/dX}\propto\frac{E_i}{\rho}=\frac{E_i}{\rho t^3}t^3,
\end{equation}
and
\begin{equation} \label{t_e propto}
t_{e } \propto \frac{(\rho t^3)^{1/2}}{ E_i^{1/2}}.
\end{equation}
For characteristic $\beta$-decay electron energy that varies with time as $E_i\propto t^{-k}$, we have $t_e\propto (\rho t^3)^{1/2/(1-k/2)}$.



\begin{figure} 
	\includegraphics[width=\columnwidth]{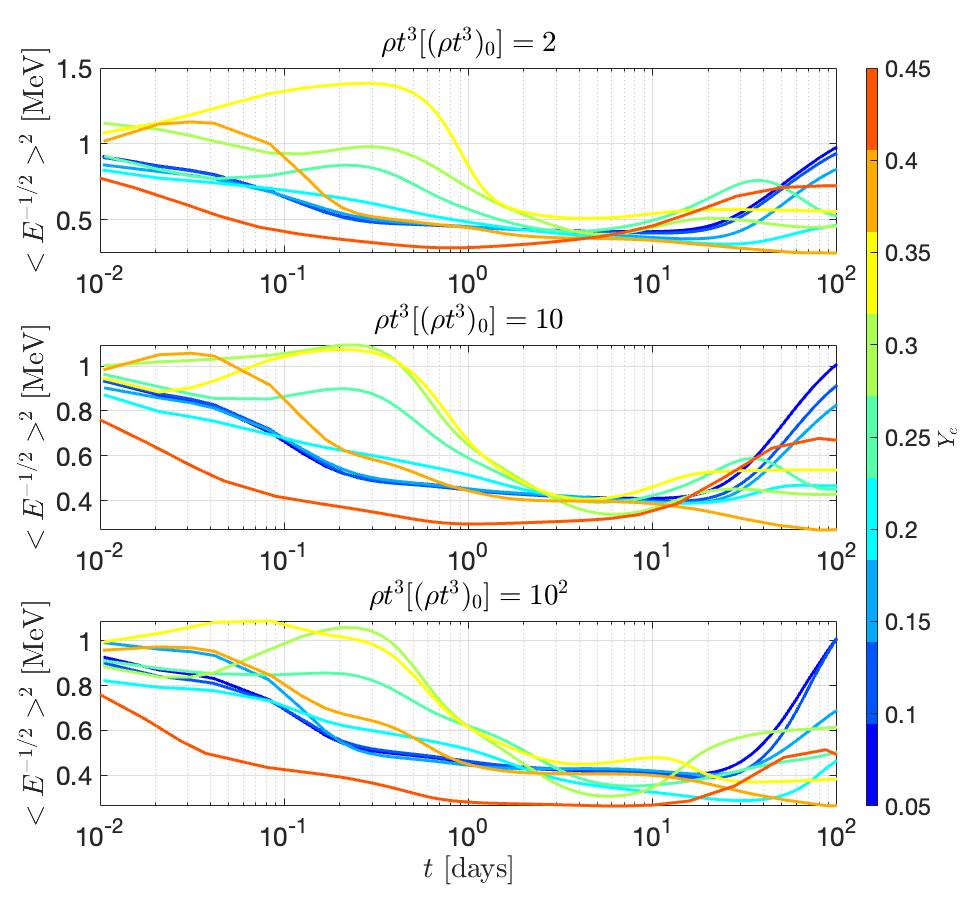}
    \caption{ Top, middle, and lower panel show $\expval{E^{-1/2}}^{-2}$ plotted vs. time for $\rho t^3/(\rho t^3)_0= 2,10,100$, respectively. Different colors correspond to different $Y_e$ values, and $s_0=20$ $k_b/\text{baryon}$ for all runs considered. We see that only for $Y_e=0.4$ does $\expval{E^{-1/2}}^{-2}$ monotonously decrease with time. Notable rises in characteristic energies occur for $Y_e = 0.05, 0.10, 0.15$ in which 3rd-peak elements are produced. This is due to the activity of "inverted decay-chains" in which slowly decaying isotopes decay to quickly decaying isotopes with higher $\expval{E^{-1/2}}^{-2}$. Their presence impacts the characteristic electron energy for nearly all nucleosynthesis calculations considered.}
    \label{beta char energy plot}
\end{figure}

Figure \ref{beta char energy plot} shows the characteristic $\beta$-decay energies vs. time for several $\rho t^3$ values (with $s_0=20$ $k_b/\text{baryon}$). Motivated by Eq. (\ref{t_e propto}), we define the characteristic energy as $\expval{E^{-1/2}}^{-2}$, rather than $\expval{E}$, where the brackets denote an average over the electron energy spectra (weighted by the electron energy). Several conclusions may be drawn from this figure. Firstly, only for $Y_e=0.4$ does the characteristic energy monotonously decrease with time up until $t \sim 100$ days. This is consistent with Figure \ref{t_e first plot}, where only the $Y_e=0.4$ curve exhibits a logarithmic slope (of the dependence of $t_e$ on $\rho t^3$) that is $>1/2$ at late times. For $Y_e=0.05,0.1,0.15$, we see a sharp rise in characteristic energy around $t \sim 20$ days. For these $Y_e$ values, we find that the radioactive heating at $t\gtrsim 30$ days becomes dominated by the following decay-chain
\begin{equation} \label{inverted reaction}
    ^{194}\text{Os} \xrightarrow[\expval{E}=0.03 \text{MeV}]{t_{1/2} \approx 6 \text{ yr}} {^{194}\text{Ir}}
    \xrightarrow[\expval{E}=1.09 \text{MeV}]{t_{1/2} \approx 20 \text{ hr}} {^{194}\text{Pt}}.
\end{equation}
This is an example of an "inverted decay-chain" in which a slowly decaying isotope decays to a rapidly decaying isotope with higher energy release. We find that at these times, other inverted decay-chains are also active and contribute significantly to the heating. These include $A = 140, 132, 106$ and others, depending on $Y_e$. We note that these occur only for even values of $A$, and may be understood using the semi-empirical mass formula \citep{weizsacker_zur_1935}. The pairing term in the formula creates two mass parabolas for a given $A$: one for even-even and another for odd-odd nuclei. As the nucleus decays, it transitions between these parabolas, resulting in uneven energy releases until reaching stability.  Overall, there are 40 inverted decay-chains for which a slowly decaying isotopes decays into a rapidly decaying isotopes with half-life $>10^2 \ t_{1/2}$ of its parent isotope, 26 of these are for decay chains with $A\geq 90$.


In Region I ($Y_e<0.22$ \& $\forall s_0$, see Table \ref{t_e table}), third peak $r-$process elements are robustly produced. Therefore, the inverted decay chains for $A=132, 140, 194$ are active, leading to an increase in late-time characteristic energy release and, thus, a lower value of the exponent, $a_2 = 0.37$. In Region II, the situation is more subtle. We note that in NSE, the entropy $s_0$ scales monotonically with the photon-to-baryon ratio $\phi$. The abundance of an isotope ($Z,A$) in NSE is proportional to $\sim \phi^{1-A}$ and thus NSE with high entropy favors more free neutrons and less heavy-nuclei \citep{meyer_entropy_1993}. As the ejecta expand and exit NSE, the many free neutrons are captured on the few seed nuclei, leading to heavy-element production. Thus for $Y_e = 0.25, 0.3$, third peak elements are produced, albeit less than in Region I by a factor $\sim10^{-3}$, despite the ejecta being relatively neutron-poor. The inverted decay chains for $A=132, 140, 194$ are similarly active. For $Y_e = 0.35, 0.4$, $r$-process elements are produced up to the second peak (up to $A\approx 125$). Despite the high entropy, the ejecta is too neutron-poor to create second and third-peak $r$-process elements. For these values of $Y_e$ in Region II, the inverted decay chains for $A=106, 125$ dominate the heating from $t\sim 10$ days, leading to an increase in characteristic initial electron energy. For these same high $Y_e$ values but for low values of $s_0$ (Region III), isotopes up to $A\approx 90$ are synthesized, and thus, none of the significant inverted decay chains are active. In Region III, the characteristic initial electron energy, for the most part, either stays constant or slightly declines over time, yielding $a_2 = 1/2$ as is expected from Eq. (\ref{t_e propto}).

We conclude that inverted $\beta$-decay chains are not insignificant anomalies that can be disregarded when considering characteristic energies of $\beta$-decay electrons. Their presence in nearly all the nucleosynthesis runs considered, especially those with significant heavy element production, invalidates the assumption that the characteristic electron energy steadily declines with time.

\subsection{Analytic approximation for $\qdep$} 
\label{qdep analytic description}

In this section we provide an interpolating function $\qint$ that describes the electron energy deposition rate both at early ($t<t_e$) and late ($t>t_e$) times, for different values of $\{ \rho t^3, Y_e, s_0 \}$. Since $\qdep \sim t^{-2.85}$ at late times regardless of $\qdot$ (see \citet{waxman_late-time_2019}), we provide an interpolation for $\qdep$ rather than for the fractional deposition, $\qdep/\qdot$. We suggest the following interpolation: 
\begin{equation}
\begin{split}
& \dot{Q}_{\text{early}}(t)  \equiv \qdot(t) \times \Bigg( 1 - \exp \Big[ - \Big( \frac{t}{t_e} \Big)^{-n_1} \Big] \Bigg), \\
& \dot{Q}_{\text{late}}(t) \equiv \dot{Q}_{\text{early}}(t=t_d) \times \Bigg( \Big( \frac{t}{t_d} \Big)^{-15} + \Big( \frac{t}{t_d} \Big) ^{-3 n_2}  \Bigg)^{1/3},  \\ 
& \dot{Q}_{\text{dep}}^{\text{int}}(t) = \big( \dot{Q}_{\text{early}}^m + \dot{Q}_{\text{late}}^m \big) ^{1/m}.
\end{split}
\label{q interp}
\end{equation}
Here, $m, n_1, n_2, t_d$ are free parameters to be fitted for and $t_e$ is given by Eq. (\ref{t_e power law}). $m$ governs the transition from the early-time to late-time interpolation. $n_1$ governs the shift from full to partial deposition until $t_e$. $n_2$ governs the late-time asymptotic deposition, which should approach a $t^{-2.85}$ behavior. $t_d$ is the time at which the delayed deposition overtakes the instantaneous deposition, and should be roughly $\sim 1-2 \times t_e$. The first term in $\dot{Q}_{\text{late}}$ ensures that for $t\ll t_e$  $\dot{Q}_{\text{early}} < \dot{Q}_{\text{late}}$ such that $\qint$ interpolates correctly between the early/late-time regimes. 

We fit $\{m, n_1, n_2, t_d\}$ values for the three different regions in $\{ Y_e, s_0 \}$ as described in \S~\ref{analytical formulae section}. Our fitting employed a gradient descent over these four parameters while minimizing the cost function defined as $\text{cost} = \frac{1}{N} \sum_{i=1}^{N} \Big| \log( \qint(t_i) / \qdep(t_i)) \Big|$, where $N=40$, $t_i$ are logarithmically spaced from $t_{min}=0$ up to $t_{max} = 6t_e$. The best-fit parameters are compiled in Table \ref{t_e table}. The values obtained for $n_2$ and $t_d$ are $n_2 \approx 2.85 , t_d = 1.3 t_e$, as is expected for the correction transition to the late-time asymptotic heating.

\begin{figure} 
	\includegraphics[width=\columnwidth]{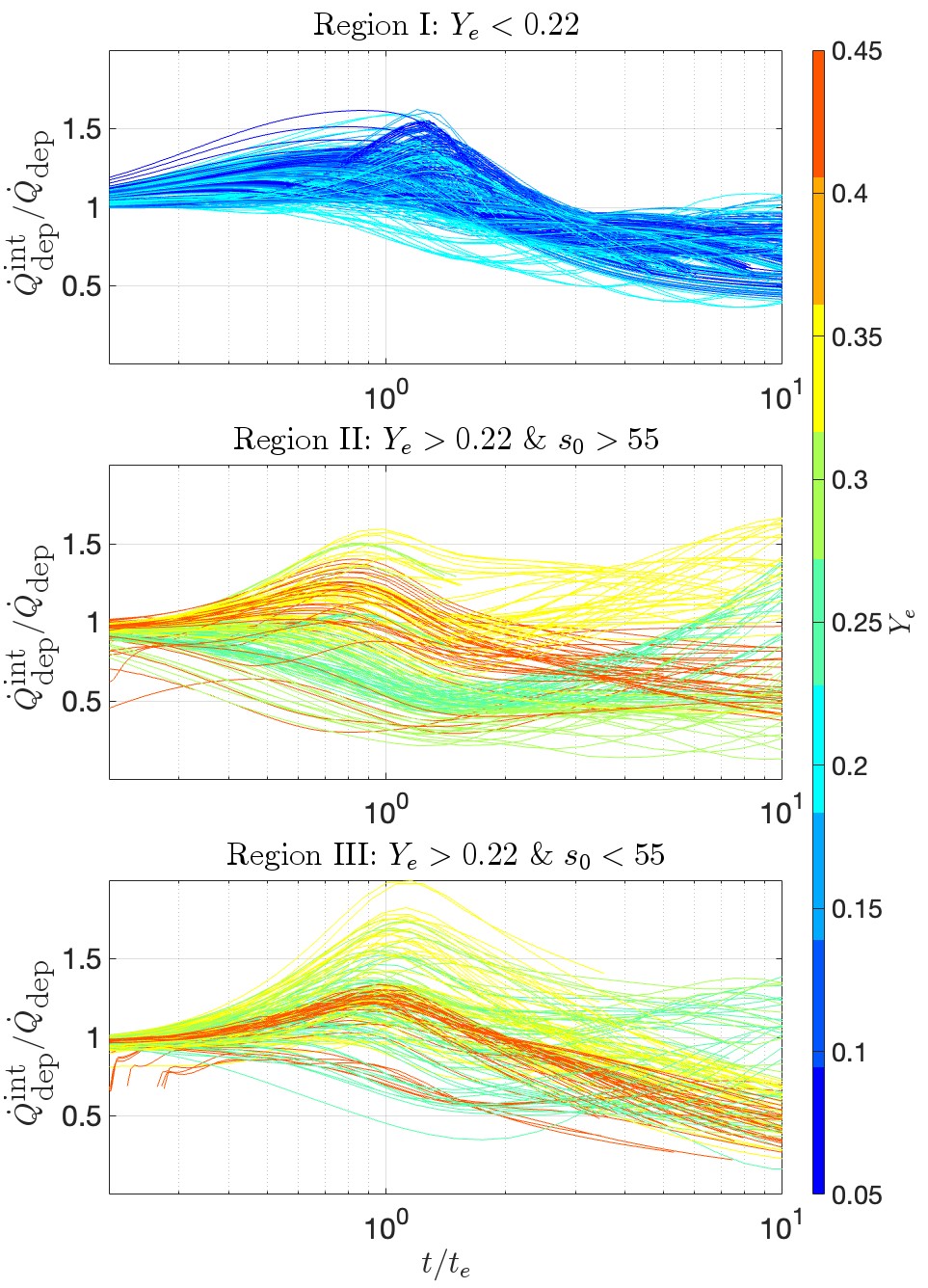}
    \caption{ $\qint / \qdep$ vs $t/t_e$ for entire parameter space of $ \{ Ye, \rho t^3 , s_0 \}$, for the three regions in parameter space as defined in Figure \ref{s2 plot}. Colors indicate different $Y_e$ values. $t_e$ is calculated according to the broken power-law fits as defined in Eq. (\ref{t_e power law}) and Table \ref{t_e table}. $\qint$ is calculated according to Eq. (\ref{q interp}) with the best-fit parameters for each region as compiled in Table \ref{t_e table}. }
    \label{q interp plot}
\end{figure}

In Figure \ref{comparing qdot, qdep, and q_interp}, we compare $\qint$ and $\qdep$ for the same sample of parameters as shown in Figure \ref{comparing energy release}.
In Figure \ref{q interp plot} we check the accuracy of the analytic description of $\qdep$ for all nucleosynthesis parameters considered. Our interpolation is most accurate in Region I, within which the nucleosynthesis yield is mostly uniform. In Regions II and III, different parameters result in different nucleosynthesis yields, and therefore different deposition curves, which reduces the accuracy of the fit. The quality of the fit and its spread may be gauged by considering the standard deviation of $\qdep^{int} / \qdep$ at each time, for each region. Averaging the standard deviation over logarithmic time for each region, we find that $\bar{\sigma} \approx 0.1 , 0.29 , 0.18$ for Regions I, II, and III, respectively. Overall, our interpolation is nearly accurate to within $50\%$ over $3-4$ orders-of-magnitude change in $\qdep$ for our entire parameter space. Note that Eq. (\ref{t_e power law}, \ref{q interp}) provide an analytic description that requires $\qdot (t)$ as input in order to determine the full electron deposition up to $t \sim 10 \ t_e$. 

\begin{figure*} 
	\includegraphics[width=\textwidth]{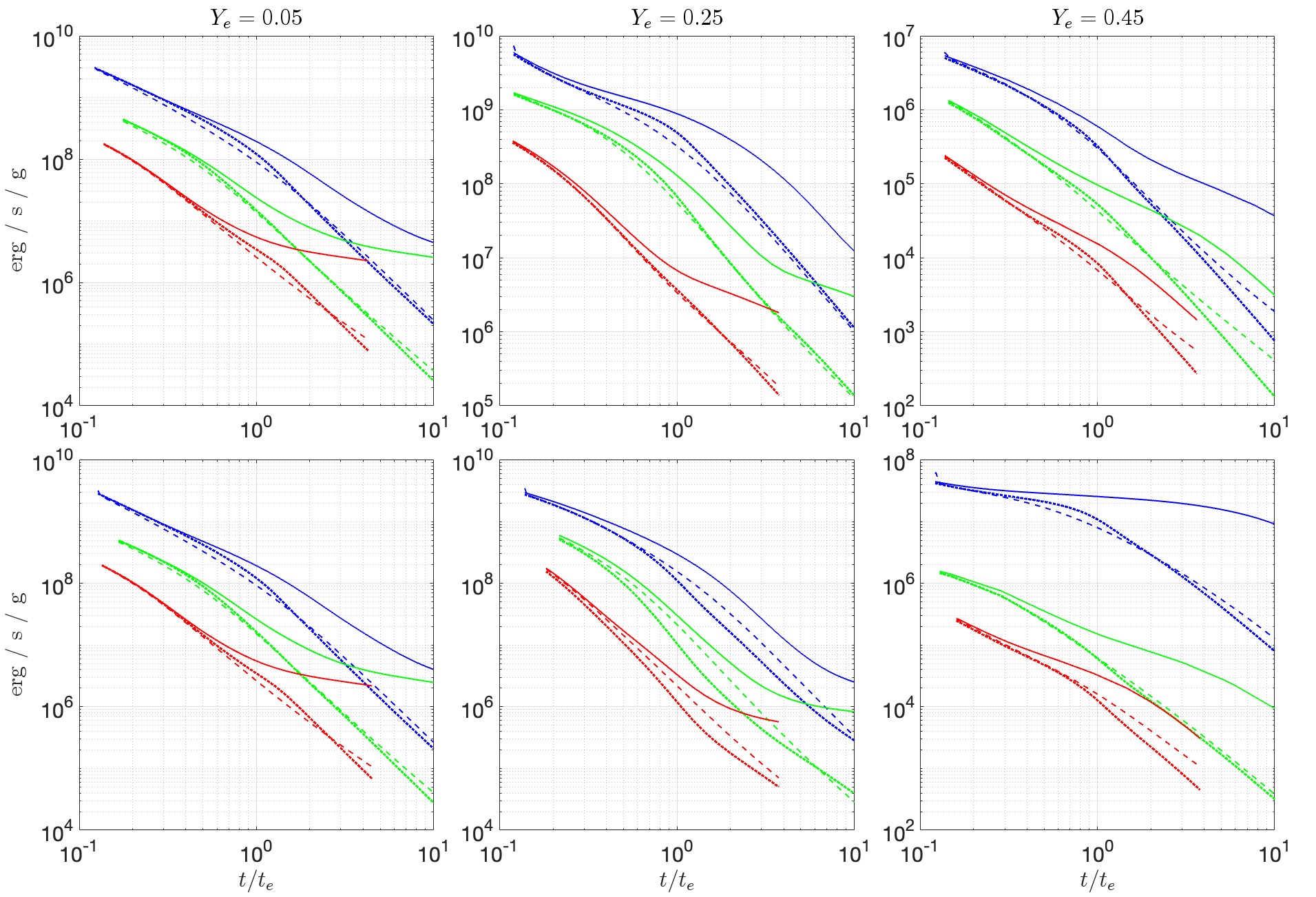}
    \caption{Electron energy release and energy deposition rates as a function of time.
    Solid and dashed lines correspond to the numerically calculated $\dot{Q}_\beta$ and $\dot{Q}_{\text{dep}}$ (Eq. (\ref{Q_dep def})) respectively. The dotted line shows our analytic approximation for $\qdep$, $\qint$ (Eq. (\ref{q interp}) and Table \ref{t_e table}). $s_0=20, (60)$ $k_b/\text{baryon}$ for top (bottom) row.
    Left, middle, and right panels correspond to $Y_e = 0.05, 0.2, 0.45$, respectively. Blue, green, and red lines correspond to $\frac{\rho t^3}{(\rho t^3)_0} = 10^{-1}, 1, 10$ respectively. }
    \label{comparing qdot, qdep, and q_interp}
\end{figure*}

\begin{table*}
\resizebox{\textwidth}{!}{%
\begin{tabular}{|c||cc||ccc||cccl||}
\hline
                                     & \multicolumn{2}{c||}{\textbf{Ejecta Parameters}}                                                   & \multicolumn{3}{c||}{\textbf{Power-Law Parameters}}                                                                   & \multicolumn{4}{c||}{\textbf{Interpolation Parameters}}                                                                                                                                                                  \\ \hline
                                     & \multicolumn{1}{c|}{$Y_e$}   & $s_0$ $[k_b/\text{baryon}]$ & \multicolumn{1}{c|}{$a_1$} & \multicolumn{1}{c|}{$a_2$} & $t_{0e}$ [days] & \multicolumn{1}{c|}{m} & \multicolumn{1}{c|}{$n_1$} & \multicolumn{1}{c|}{$n_2$} & $t_D [t_e]$ \\ \hline
\textbf{Region I}   & \multicolumn{1}{c|}{$<0.22$} & $\forall s_0$                                             & \multicolumn{1}{c|}{0.5}   & \multicolumn{1}{c|}{0.37}  & 19.5                                           & \multicolumn{1}{c|}{4.5}                             & \multicolumn{1}{c|}{1.1}                             & \multicolumn{1}{c|}{2.8}                             & 1.3                             \\ \hline
\textbf{Region II}  & \multicolumn{1}{c|}{$>0.22$} & $>55$                                                     & \multicolumn{1}{c|}{0.5}   & \multicolumn{1}{c|}{0.42}  & 19.4                                           & \multicolumn{1}{c|}{0.8}                             & \multicolumn{1}{c|}{0.5}                             & \multicolumn{1}{c|}{2.5}                             & 1.3                             \\ \hline
\textbf{Region III} & \multicolumn{1}{c|}{$>0.22$} & $<55$                                                     & \multicolumn{1}{c|}{0.5}   & \multicolumn{1}{c|}{0.5}   & 16.3                                           & \multicolumn{1}{c|}{1.5}                             & \multicolumn{1}{c|}{0.5}                             & \multicolumn{1}{c|}{2.8}                             & 1.3                             \\ \hline
\end{tabular}}
\caption{Fitted parameters for an analytic description of $t_e$ and $\qint$. Left panel: Ejecta parameters for each region in parameter space of $\{ Y_e, s_0 \}$. Middle panel: Fitted broken power-law parameters for $t_e (\rho t^3)$ as defined in Eq. (\ref{t_e power law}). Right panel: Fitted interpolation parameters as defined in Eq. (\ref{q interp}). The regions are those outlined in Figure \ref{s2 plot}. The accuracy of the interpolation is examined in Figure \ref{q interp plot}.} \label{t_e table}
\end{table*}

\subsection{Nuclear Physics Robustness Checks} \label{robustness checks}

\begin{figure} 
	\includegraphics[width=\columnwidth]{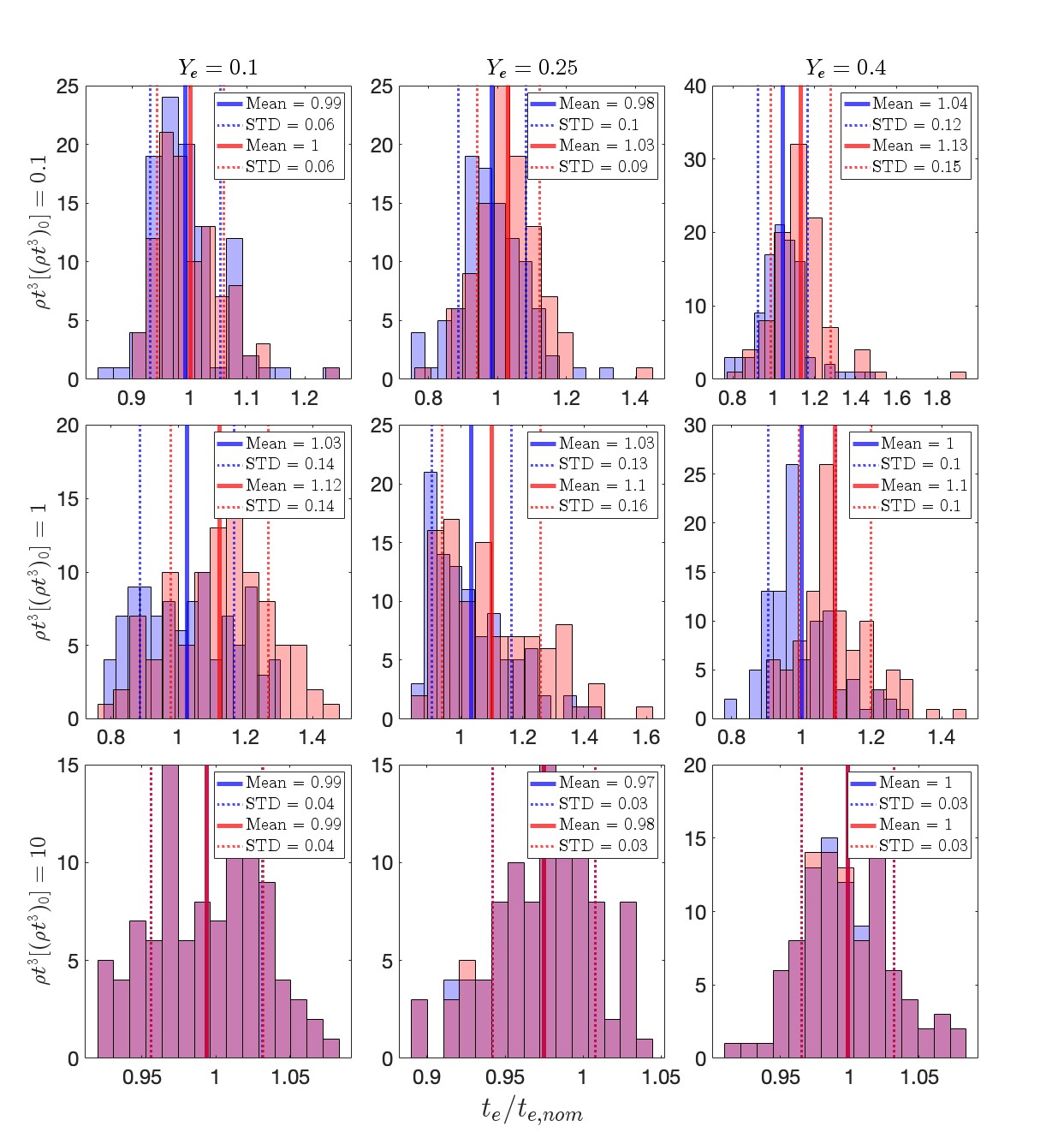}
    \caption{Counts histograms of $t_e/t_{e,\text{nom}}$, where $t_e$ is calculated 100 times for runs with different randomized reaction-rates file. $s_0 = 20 k_b/$ baryon for all runs considered. Top, middle, and bottom rows show results for $\rhoO = 10^{-1},1,10$, respectively. Left, middle, and right columns show results for $Y_e = 0.1, 0.25, 0.4$, respectively.  Blue histograms plot results using our fiducial FRDM mass-model; red histograms plot results using UNEDF1 mass-model \citep{kortelainen_nuclear_2012}. Mean is close to 1 for all, $ \sigma \approx 0.1$. }
    \label{nuclear robustness figure}
\end{figure}

Modeling $r$-process nucleosynthesis from BNS mergers requires calculating nuclear masses and nuclear reaction rates for which no experimental data are available. For $Y_e < 0.25$, different nuclear mass models may result in orders-of-magnitude differences in final composition of post 1st-peak elements ($A \gtrsim 100$) and up to an order-of-magnitude difference in radioactive energy release \citep{mumpower_impact_2016,  zhu_modeling_2021, barnes_kilonovae_2021}. In this section we study the sensitivity of our results to the nuclear physics models' uncertainties. 

To gauge the impact of these uncertainties, we reran our calculations for two different nuclear mass models together with randomized nuclear rates. We edited our local REACLIB file which contains all the rates and cross-sections that go into \textit{SkyNet}. We modify every strong or weak theoretical interaction rate $\lambda$ to
\begin{equation}
    \lambda \longrightarrow C \lambda,
\end{equation}
where $C \in [10^{-2}, 10^2]$ was randomly chosen from a uniform distribution in $\log(C)$. This amounts to changing $\sim$70,000 rates (approximately 90\% of all rates included). With our new reaction rates file, we reran $SkyNet$ for a subset of initial parameters: $\rho t^3/(\rho t^3)_0=0.1,1,10$, $s_0 = 20,70$ $k_b/\text{baryon}$, $Y_e =0.1, 0.25, 0.4$ for both FRDM and UNEDF1 \citep[][\url{http://massexplorer.frib.msu.edu},]{kortelainen_nuclear_2012} mass models. We repeat the above process 100 times and calculate $t_e$ for each run and compare the values obtained to our nominal values, $t_{e,nom}$ presented in Figures \ref{t_e first plot}. We stress that these changes lead to orders-of-magnitude differences in final composition between different runs. 

In Figure \ref{nuclear robustness figure}, we show histograms of $t_e/t_{e,nom}$ for different values of $\rho t^3$ and $Y_e$, where $t_e$ is calculated according to Eq. (\ref{t_e def}) for each \textit{SkyNet} run with randomized reaction rates using both FRDM and UNEDF1 mass models. We see that the mean falls at $\approx 1$ for all parameter values considered. The spread in $t_e$ values is small, with an average fractional standard deviation of $\sigma_{e, avg} \approx 0.12$ for $Y_e < 0.4$. For $Y_e = 0.4$, the spread is even smaller, with $\sigma_{e, avg} \approx 0.03$. 

These tests demonstrate that our results are only weakly sensitive to nuclear physics uncertainties. Different nuclear mass models and different reaction rates may lead to large variations in composition and heating rates. However, these changes do not significantly affect the inefficient thermalization timescales. In retrospect, this was to be expected - Figure \ref{t_e first plot} shows that $t_e$ is relatively insensitive to the $Y_e$ value and thus to the detailed electron energy spectra and ejecta composition.

\section{Conclusions} \label{conclusions}

We calculated the time-dependent energy deposition rate per unit mass, $\qdep$ (see Eq. (\ref{Q_dep def})), by electrons (and positrons) emitted by $\beta$-decays in homologously expanding ejecta produced in BNS mergers for a wide range of ejecta parameters, $\{\rho t^3,s_0, Y_e\}$. Using detailed numeric nucleosynthesis calculations, we obtained the time-dependent composition, stopping power, and $\beta$-decay energy spectra, and followed the evolution of the electron energy distribution, assuming that the electrons are confined by magnetic fields to the fluid element within which they were produced (see Eq. (\ref{electron energy evolution})).  

The deposition efficiency, the ratio $f_{\rm inst}$ between the rates of instantaneous electron energy deposition and energy production rates, Eq.~(\ref{f_dep}), depends mainly on $\rho t^3$ and only weakly on $\{s_0, Y_e\}$. $t_{e}$, the time at which $f_{\rm inst}$ drops to $1-e^{-1}$, is well described by a broken power-law, $t_{e } = t_{0e}\times \Big( \frac{\rho t^3}{0.5 (\rho t^3)_0} \Big)^{a_{1,2}}$, see Eq.~(\ref{t_e power law}) and Figure \ref{t_e 3 regions plot}, with $t_{0e}(s_0, Y_e) \approx 17$~days, and $0.4\le a(s_0, Y_e)\le0.5$ (see Table \ref{t_e table} for exact values). The accuracy of the analytic approximation is within $\sim 20 \%$. 

The result $a_{1,2} \le 0.5$ reflects the fact that contrary to naive expectations, the characteristic energy of the $\beta$-decay electron spectrum does not decrease with time (see Eq.~(\ref{t_e propto})). This is largely due to "inverted decay chains" in which a slowly-decaying isotope decays to a rapidly-decaying isotope with higher end-point energy. A detailed discussion of the relevant decay chains at different \{$Y_e,s_0$\} values is given in \S~\ref{char energy release electrons}. We find that inverted decay chains play a significant role at $t \gtrsim 10$~days, affecting the thermalization timescale at all relevant \{$Y_e,s_0$\} values. 

Using our analytic results for $t_{e}$, we provided an analytic description of the energy deposition of $\beta$-decay electrons (and positrons), $\qint (t)$ given by Eq.~(\ref{q interp}) and Table \ref{t_e table}. Our formulae provide an analytic description of the thermalization efficiency, which we find to be only weakly sensitive to $\{s_0, Y_e\}$ (see \S~\ref{analytical formulae section}, \ref{char energy release electrons}, and Figures \ref{t_e first plot} , \ref{s2 plot} , \ref{t_e 3 regions plot}) and nuclear physics uncertainties (see \S~\ref{robustness checks}, Figure \ref{nuclear robustness figure}). This enables one to determine $\qint (t)$ for a given total radioactive energy release rate, $\qdot(t)$. Our analytic description reproduces the numerical results to better than $\approx 50\%$ (typically $<30\%$) accuracy up to $t \lesssim 10 \ t_e$, over a $3-4$ orders-of-magnitude deposition rate decrease (see \S~\ref{qdep analytic description}, Figures \ref{comparing qdot, qdep, and q_interp} , \ref{q interp plot}). 

In contrast to previous works that compile tables of parameterized fits for the thermalization of specific combinations of ejecta mass and velocity in a specific velocity structure (e.g. \cite{barnes_radioactivity_2016}), we provide a simple analytic description that can be utilized for a wide range of ejecta parameters and density profiles. Our results may thus be easily incorporated in calculations of kilonovae light curves (with general composition, density and velocity structures), eliminating the need to follow the time-dependent electron spectra numerically. 

The weak dependence of the deposition efficiency on $\{Y_e,s_0\}$ and on nuclear physics uncertainties implies that identifying $t_e$ in the bolometric kilonova light curve will constrain the (properly averaged) ejecta $\rho t^3$.

\section*{Acknowledgements}
    BS would like to thank Tal Wasserman for helpful discussions.

\section*{Data Availability}

Numerical results are available upon request.



\bibliographystyle{mnras}
\bibliography{elec_dep_refs}



\bsp	
\label{lastpage}
\end{document}